\documentclass[aps,prl,superscriptaddress,twocolumn,showpacs]{revtex4-1}
\usepackage[utf8]{inputenc}
\usepackage[american,british]{babel}
\usepackage[T1]{fontenc}
\usepackage[pdftex]{graphicx}
\usepackage{xcolor}
\usepackage{dcolumn}
\usepackage{bm}
\usepackage{amsmath,amsthm,amssymb,mathrsfs,mathtools,amsfonts,braket}
\usepackage{verbatim}
\usepackage{epsfig}
\usepackage{color}
\usepackage{xfrac}
\usepackage{geometry}
\usepackage{hyperref}
\usepackage{bbold}
\usepackage{tensor}
\usepackage{ulem}

\renewcommand{\sim}{\thicksim}

\usepackage{hyperref}
\hypersetup{
    bookmarks=false,         
    unicode=false,          
    pdftoolbar=false,        
    pdfmenubar=true,        
    pdffitwindow=false,     
    pdfstartview={FitH},    
    pdftitle={},    
    pdfauthor={Authors},     
    pdfsubject={},   
    pdfcreator={},   
    pdfproducer={}, 
    pdfnewwindow=true,      
    colorlinks=true,       
    linkcolor=black,          
    citecolor=blue,        
    filecolor=magenta,      
    urlcolor=blue           
}

\begin{document}

\preprint{APS/123-QED}

\title{Overcoming the entanglement barrier in quantum many-body dynamics \\ via space-time duality}

\author{Alessio Lerose}
\affiliation{Department of Theoretical Physics,
University of Geneva, Quai Ernest-Ansermet 30,
1205 Geneva, Switzerland}

\author{Michael  Sonner}
\affiliation{Department of Theoretical Physics,
University of Geneva, Quai Ernest-Ansermet 30,
1205 Geneva, Switzerland}

\author{Dmitry A. Abanin}
\affiliation{Department of Theoretical Physics,
University of Geneva, Quai Ernest-Ansermet 30,
1205 Geneva, Switzerland}


\begin{abstract}

Describing non-equilibrium properties of quantum many-body systems is challenging due to high entanglement in the 
wavefunction.
We describe evolution of local observables via the influence matrix (IM), which encodes the  effects of a many-body system as an environment for  local subsystems.
Recent works found that in many dynamical regimes the IM of an infinite system has low {\it temporal entanglement} and can be efficiently represented as a matrix-product state (MPS).
Yet, direct iterative constructions of the IM
encounter highly entangled intermediate  states -- a temporal entanglement barrier (TEB).
We argue that TEB is ubiquitous, 
and elucidate its physical origin via a semiclassical quasiparticle picture that exactly captures the behavior of integrable spin chains. Further, 
we show that a TEB also arises in chaotic spin chains, which lack well-defined quasiparticles. Based on these insights, we formulate an alternative {\it light-cone growth algorithm}, which provably avoids TEB, thus providing an efficient construction of the thermodynamic-limit IM  as a MPS.
This work uncovers the origin of the efficiency of the IM approach for thermalization and transport. 
\end{abstract}

\maketitle

Describing quantum many-body dynamics is challenging, as the exponential complexity of general many-body wavefunctions $\ket{\Psi}$ limits the reach of numerical simulations. While area-law entangled ground states can be compactly represented using tensor networks~\cite{schollwoeckrev}, out-of-equilibrium states $\ket{\Psi(t)}$ typically exhibit an exponential blowup of complexity, associated with the dynamical growth of quantum entanglement
~\cite{Calabrese_2005,Kim13,NahumPRXEntanglement}, which precludes their efficient parametrization. This poses fundamental challenges for the computation of properties such as  transport character and coefficients.

Several approaches have been recently proposed to capture quantum many-body dynamics  beyond this ``entanglement wall"~\cite{leviatan2017quantum,kloss2018tdvp,white2018dmt,surace2019simulating,ye2020emergenthydro,krumnow2019towards,rakovszky2020dissipation,Karrasch_2013,hauschild2018finding}.
Their common strategy is to modify the time-evolution update rule of the relevant object
(many-body wavefunction~\cite{leviatan2017quantum,kloss2018tdvp,krumnow2019towards}, density matrix~\cite{white2018dmt,surace2019simulating,ye2020emergenthydro,Karrasch_2013,hauschild2018finding},
Heisenberg-picture observable~\cite{rakovszky2020dissipation}) by introducing a truncation or damping
scheme for the non-local quantum correlations, at the same time preserving local correlations, i.e. the physically relevant information.
While these approaches successfully describe diffusive dynamics in certain quantum spin chains,
their 
general applicability 
remains unclear.

Working with the
\textit{instantaneous} quantum state of the full system (e.g., of $|\Psi(t)\rangle$) 
can be disadvantageous, since it is difficult to discern the correlations
between a subsystem $\mathcal{S}$ and its complement $\mathcal{E}$ which do affect the later evolution of
$\mathcal{S}$, from those which do not.
Inspired by the framework of open
quantum systems~\cite{FeynmanVernon,LeggettRMP},  we recently formulated the influence matrix (IM) approach to study dynamics in quantum circuits~\cite{lerose2020influence,sonner2021aop}.
In this approach one compresses the actual many-body environment $\mathcal{E}$ to an effective environment $\widetilde{\mathcal{E}}$ with a smaller dimension $\chi$, exerting nearly identical dynamical influence on $\mathcal{S}$.
 The influence functional~\cite{FeynmanVernon} of this compressed  environment  then takes the form of a matrix-product state (MPS) in the temporal domain, with a bond dimension $\chi$, encoding temporal correlations and memory effects (i.e., non-Markovianity). 
 The advantage of this approach is that the quantum information that flows from $\mathcal{S}$ to $\mathcal{E}$, but does not affect later dynamics of $\mathcal{S}$, is neglected, allowing to reduce $\chi$ to the minimum size necessary to accurately reproduce the evolution of $\mathcal{S}$.

The IM's intrinsic complexity is characterized by the scaling of $\chi$ with evolution time.
By analogy with conventional many-body wave functions in space, this complexity is related to the {\it temporal entanglement} (TE) entropy of the IM~\cite{lerose2020influence,Banuls09}.
Recently found exact solution and numerical studies~\cite{lerose2020influence,lerose2021integrable,Piroli2020exact,Koblas20,Klobas2021exact,giudice2021temporal,Sonner20CharacterizingMBL} indicate that TE is low in several distinct regimes, suggesting computational advantage over conventional methods.
%
However, $\chi$ determines  the full computational complexity of the IM approach
 only if  the compression procedure of the original many-body environment, i.e., the construction of the optimal MPS form of the IM, can itself be done efficiently. This has been recognized as a distinct challenge~\cite{Chan21,sonner2021aop}.
 For one-dimensional systems, the direct iterative contraction of the IM tensor network -- akin to the  ``transverse contraction and folding'' method previously proposed in Refs.~\cite{Banuls09,muller2012tensor} -- was empirically observed to lead to
a  {\it temporal entanglement barrier} (TEB)~\cite{Chan21,sonner2021aop}:  states with very high TE were encountered in intermediate steps of the computation, 
even though  
IM itself had low TE.

In this work we address the latter problem, laying the theoretical foundations of  the efficiency of the IM approach.
Specifically, we elucidate  the physical origin of TEB 
and identify a crucial modification of the \textit{na\"{i}ve} compression strategy for one-dimensional systems that allows us to circumvent TEB. Our findings imply that computation of the dynamics of local observables is only limited by the  compressed dimension $\chi$ achievable for the IM of the given system. {We illustrate the performance of the new algorithm by studying quench dynamics of a non-integrable spin chain, finding that it outperforms existing schemes.}



\begin{figure}
    \centering
    \includegraphics[width=0.46\textwidth]{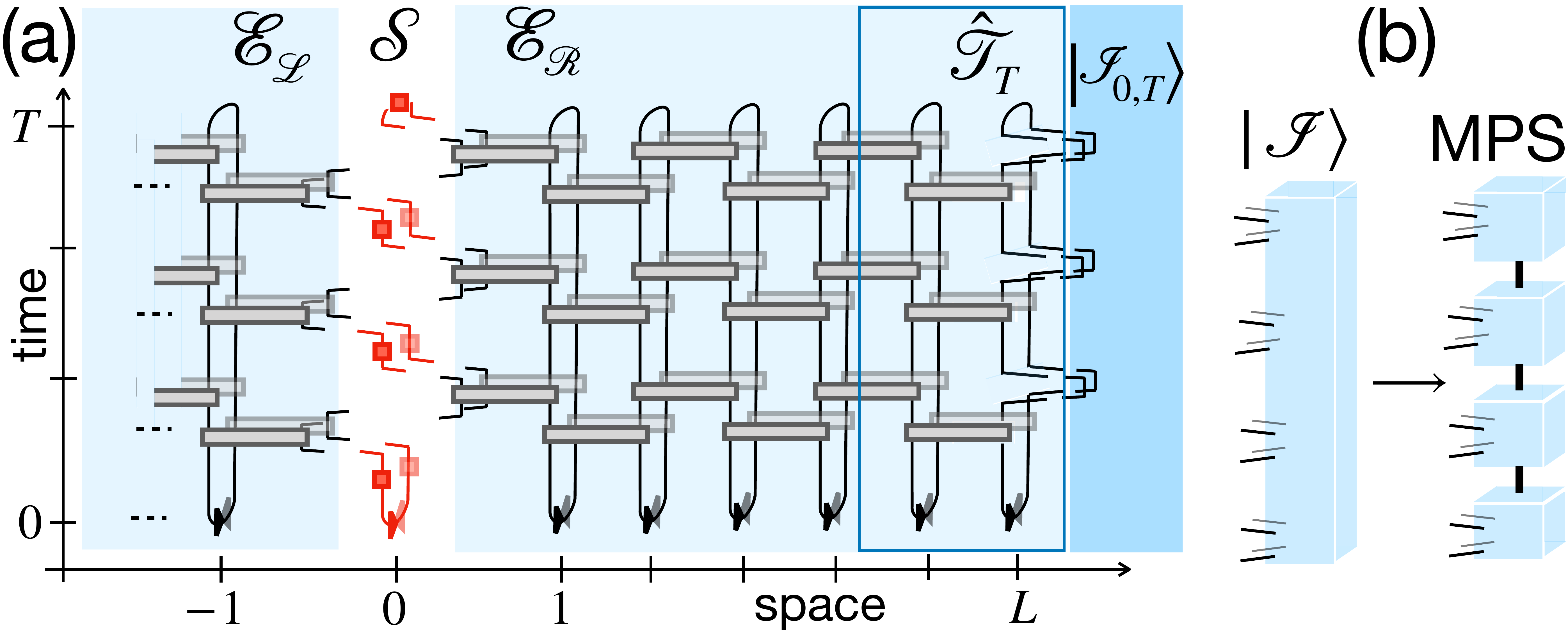}
    \caption{
    {\it (a)}:
    IM approach for a quantum spin chain. Evolution of ``ket'' and ``bra'' components of the initial density matrix (here spatially uncorrelated, black, bottom) is generated by two conjugated  unitary circuits (foreground and background  sheets of gray ``bricks''). Red squares
    denote operators  acting only on $\mathcal{S}$ (here a single spin).
    Their time-ordered correlators are influenced by the spins in the left and right complements $\mathcal{E}_\mathscr{L,R}$ of $\mathcal{S}$, traced out at final time~$T$.
    The blue-shaded tensors are the two associated IMs. 
    The OBC IM $|\mathscr{I}_{0,T}\rangle$ and the dual transfer matrix $\hat {\mathcal{T}}_T$  are highlighted.
    {\it (b)}: MPS representation of an IM.
    }
    \label{fig_IM}
\end{figure}

\textit{\bf Temporal entanglement barrier.}
We consider spin-$1/2$ chains with local interactions, initialized in a factorized state, and take $\mathcal{S}$ to consist of one or more adjacent spins.
Within the IM approach, the degrees of freedom in the complement $\mathcal{E}$ are traced out, leaving an effective open quantum system evolution for~$\mathcal{S}$.
The influence of $\mathcal{E}$ on $\mathcal{S}$ is encoded in non-local-in-time ``self-interactions'' 
of $\mathcal{S}$~\cite{FeynmanVernon}.
Concretely, we express dynamics of $\mathcal{S}$ via a 
Keldysh path integral over trajectories of spins in $\mathcal{S}$ and $\mathcal{E}$.
Path integration over the latter results in a functional $\mathscr{I}$ of the trajectory of $\mathcal{S}$.
This can be obtained by representing 
time-evolution as a unitary circuit~\cite{vidaltebd}, and viewing the IM
as the contraction of the associated tensor network [blue shaded region in Fig.~\ref{fig_IM}{(a)}]~\cite{lerose2020influence}.
%
%
The IM can be viewed as a fictitious one-dimensional wavefunction on the
Schwinger-Keldysh contour $|\mathscr{I}\rangle$ and thus compressed as MPS [Fig.~\ref{fig_IM}{(b)}]~\cite{lerose2020influence,Chan21}.

Previous works~\cite{Banuls09,lerose2020influence,Chan21,sonner2021aop} found the IM of an infinite chain by repeatedly applying a dual transfer matrix $\hat {\mathcal{T}}$ to some initial vector,  representing the intermediate wavefunctions as MPSs and truncating them to a bond dimension cutoff $\chi$.
Making reference to Fig.~\ref{fig_IM}{(a)}, we consider the IM $|\mathscr{I}_{L,T}\rangle$ of the right environment $ \mathcal{E}_{\mathscr{R}}$ ($\equiv\mathcal{E}$ henceforth), consisting of $L$ spins, for evolution time $T$.
$\hat {\mathcal{T}}_T$ acts from right to left, including two extra sites in the environment at each application. The $L/2$ iterations of $\hat {\mathcal{T}}_T$ start from a boundary IM $|\mathscr{I}_{0,T}\rangle$.
In Fig.~\ref{fig_IM}{(a)}, $|\mathscr{I}_{0,T}\rangle$ is the IM of an empty environment due to open boundary conditions (OBC)  -- a product state of local Bell pairs.
As the exact IM  $|\mathscr{I}_{\ell,T}\rangle=\hat {\mathcal{T}}_T^{\ell/2}|\mathscr{I}_{0,T}\rangle$ ``evolves'' in space, 
iterations converge to the infinite-system IM $|\mathscr{I}_{\infty,T}\rangle$ for $L \ge  v_{\text{max}} t$, where $v_{\text{max}}$ is the Lieb-Robinson velocity~\cite{LiebRobinson} of the model, regardless of the choice of $|\mathscr{I}_{0,T}\rangle$.
Truncation with cutoff $\chi$ can however affect the convergence.
References~\cite{Chan21,sonner2021aop}, in fact, observed that the TE of intermediate IMs $|\mathscr{I}_{\ell,T}\rangle$ is non-monotonic in $\ell$, reaching values much higher than that of $|\mathscr{I}_{\infty,T}\rangle$ -- the behavior that we refer to as TEB.
Intuitively, TEB can be traced back to the fact that $\mathcal{T}_T^{\ell/2}$ enables one to compute local evolution with  finite chain environments of length~$\ell$. 
Such environments generally allow back-flow of quantum information injected from $\mathcal{S}$, typically resulting in extensive TE.

\begin{figure*}
    \centering
    \includegraphics[height=0.16\textwidth]{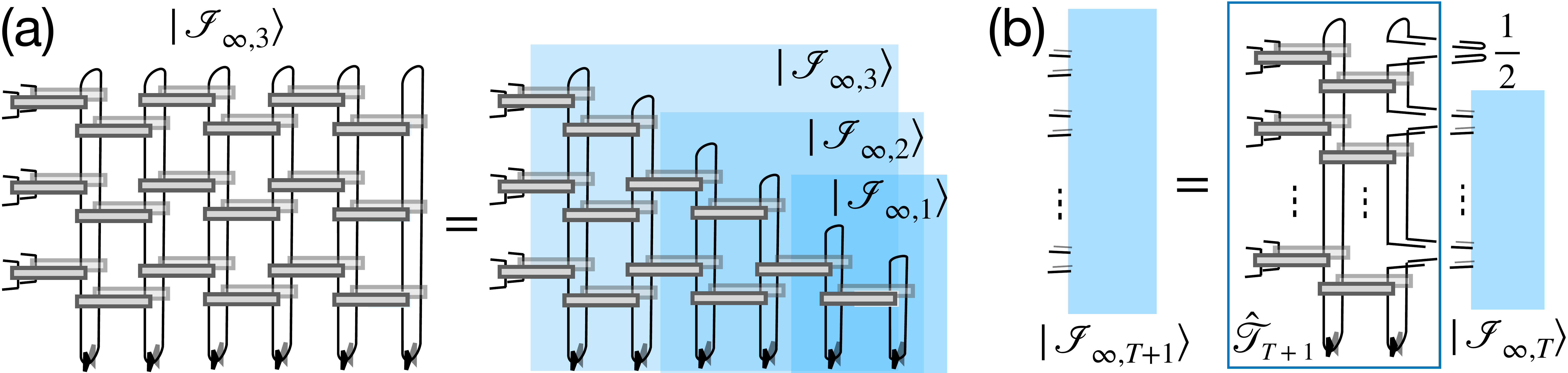}
    \hspace{0.1cm}
    \includegraphics[height=0.16\textwidth]{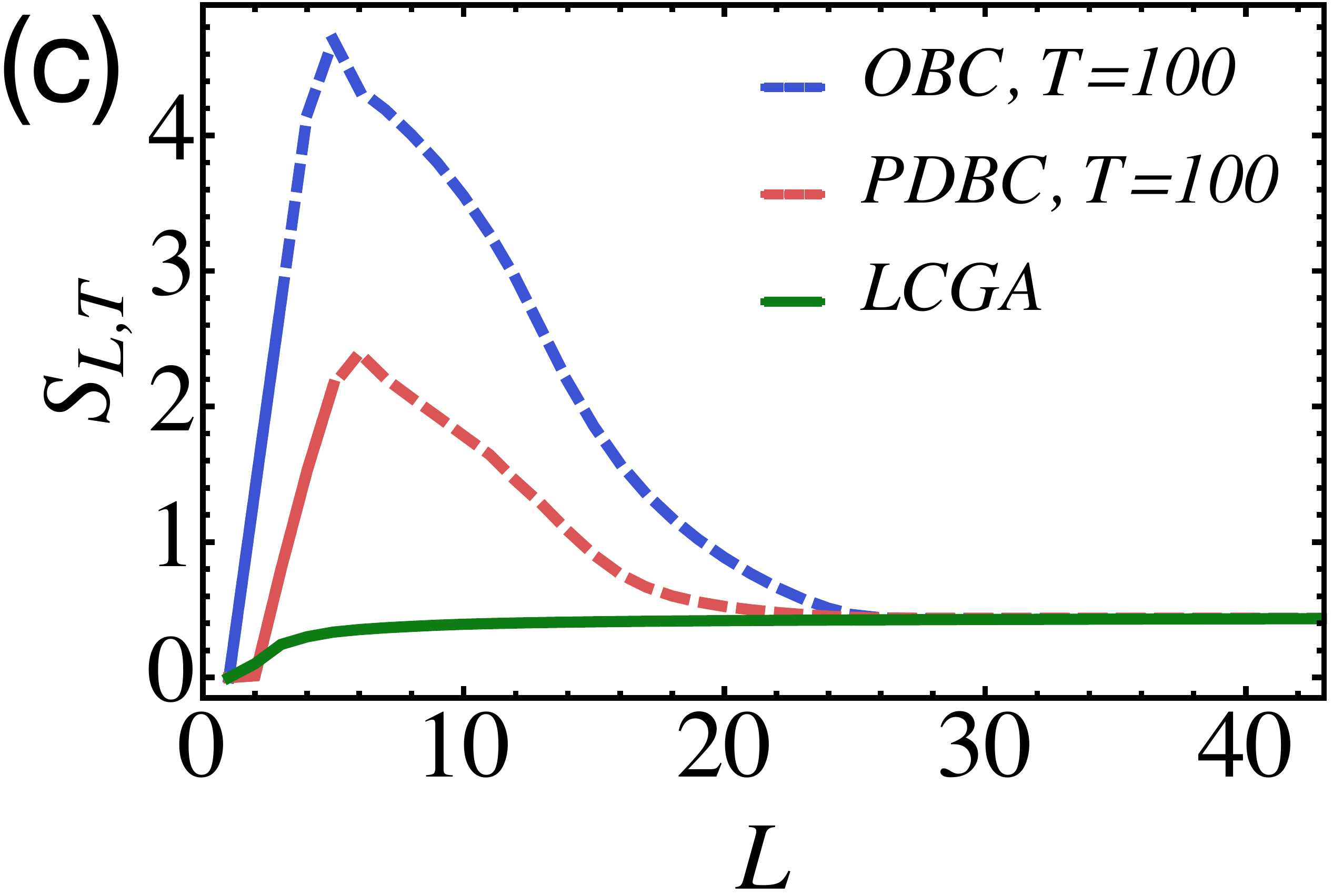}
    \caption{
    {\it (a)}:   
    Intermediate LCGA steps  produce the sequence of bulk IMs for increasing $T$ (blue-shaded boxes). 
     {\it (b)}:
    Iterative LCGA step.
     {\it (c)}: Comparison of TE vs iterative step of $\hat{\mathcal{T}}$-iterations with OBC, PDBC and LCGA, for the KIC in Eq.~\eqref{eq_KIC} with $g=0.685$, $J=0.31$, $h=0.2$. Dashing indicates that data are not converged for $\chi=128$. 
    }
    \label{fig_lc}
\end{figure*}

\begin{figure}
    \centering
     \includegraphics[width=0.36\textwidth]{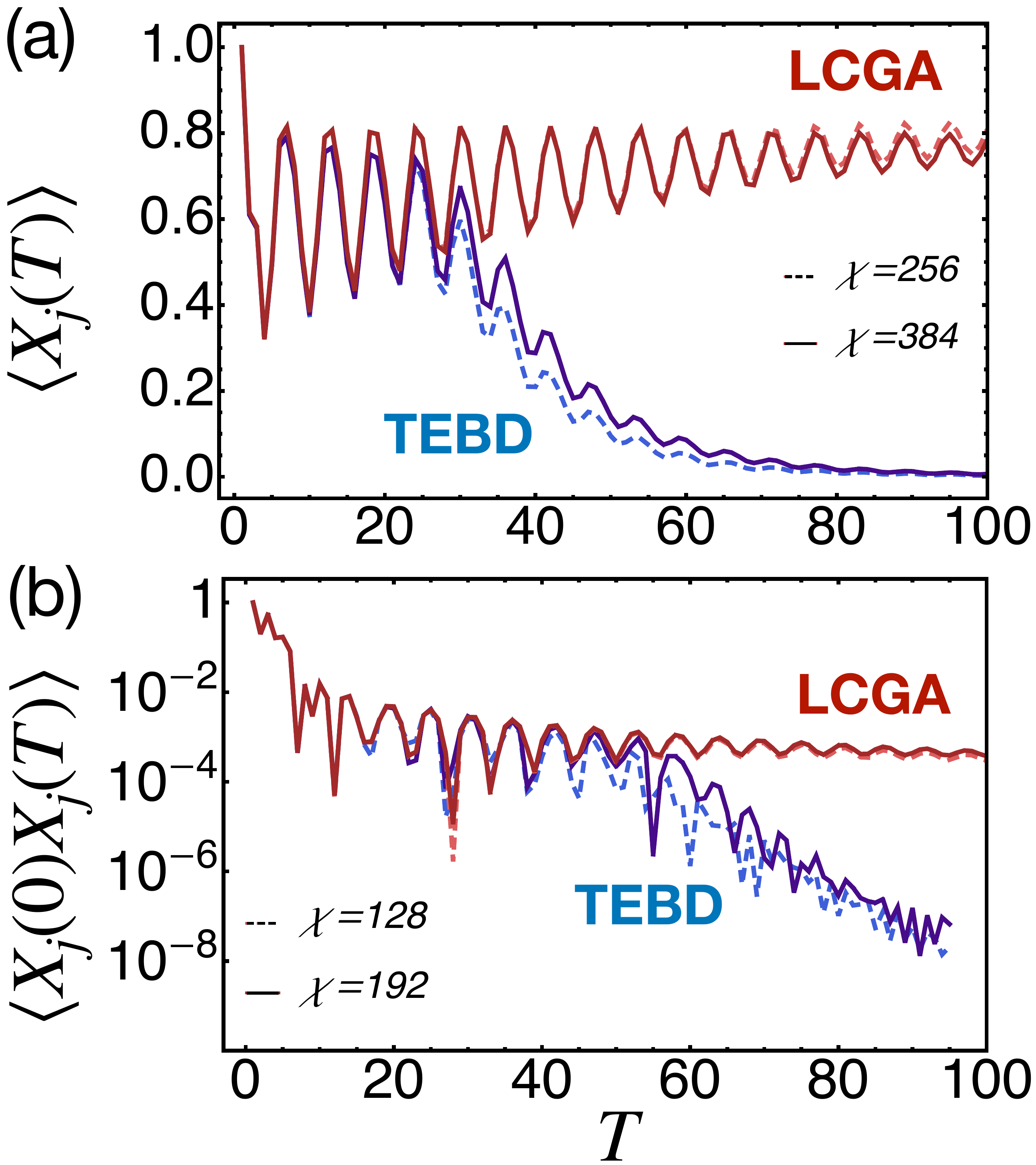}
    \caption{
     Comparison between simulation of long-time dynamics of the KIC in Eq.~\eqref{eq_KIC} with $g=0.685$, $J=0.31$, $h=0.2$, using LCGA (red) and TEBD (blue lines), with the same set of bond dimension cutoffs $\chi$ for the two methods.
     {\it (a)}: Evolution of $X$ polarization starting from the initial  state fully polarized along $X$. Note that heating is dynamically slow in this regime due to incommensurate $g$~\cite{DeRoeckVerreet19}.
     {\it (b)}: Temporal autocorrelator of the local $X$ polarization at infinite temperature.
    }
    \label{fig_benchmark}
\end{figure}

\textit{\bf Light-cone growth algorithm.}
To circumvent TEB, we  introduce a modified algorithm dubbed \textit{light-cone growth algorithm} (LCGA) which avoids finite environments by combining
contractions in the spatial and temporal directions (see Fig.~\ref{fig_lc}). 
The circuit defining the infinite-system IM is equivalent to a finite ``triangular'' network,
obtained by erasing the unitary gates located beyond the light ray emanating
backward in time from $\mathcal{S}$. By causality, such gates cannot produce any
influence on $\mathcal{S}$ -- a manifestation of the strict light-cone effect of
circuit dynamics.
We now contract  
this reduced network in the space direction.
At the $n$-th step, the computation produces
the {\it infinite-system} IM corresponding to evolution time $T=n$, $|\mathscr{I}_{2n,n}\rangle\equiv|\mathscr{I}_{\infty,n}\rangle$.
The following one $|\mathscr{I}_{2n+2,n+1}\rangle\equiv|\mathscr{I}_{\infty,n+1}\rangle$ for time $T=n+1$ is obtained by applying the dual transfer matrix $\hat{\mathcal{T}}_{n+1}$ to the previous IM augmented by identities on two extra sites, i.e.  $|\mathscr{I}_{2n,n}\rangle\otimes |\mathbb{1}_2\rangle\otimes \frac 1 2  |\mathbb{1}_2\rangle$, as illustrated in Fig.~\ref{fig_lc}(b).

For translationally invariant chains,
this algorithm circumvents TEB, since $S_{\infty,T}$ and the necessary bond dimension $\chi_{\infty,T}$ increase monotonically with $T$.
Intuitively, this is because all the information encoded in $|\mathscr{I}_{\infty,T}\rangle$ is contained in $|\mathscr{I}_{\infty,T+t}\rangle$.
Monotonicity of the bond dimension, i.e., $\chi_{\infty,T} \le \chi_{\infty,T+t}$ (for any given bond) can be proven by noting that
$|\mathscr{I}_{\infty,T}\rangle$ is obtained from $|\mathscr{I}_{\infty,T+t}\rangle$ by projecting its last $t$ pairs of legs onto $\bigotimes_{\tau=T+1}^{T+t}\frac{1}{2}\langle\mathbb{1}|\otimes\langle\mathbb{1}|$, which follows from erasing the last $t$ layers of gates using unitarity [cf. Fig.~\ref{fig_lc}(a)].
By construction, the bond dimension at any bond $(\tau,\tau+1)$ cannot increase upon performing this local projection operation.



To demonstrate that LCGA resolves the TEB issue, 
in Fig.~\ref{fig_lc}(c) we compare three different computations of the bulk IM for the infinite-temperature kicked Ising chain (KIC), described by the Floquet operator
\begin{equation}
\label{eq_KIC}
    U =  e^{i g \sum_j X_j}  e^{i J \sum_j Z_j Z_{j+1}}
    e^{i h \sum_j Z_j}
\end{equation}
with $X_j$, $Z_j$ Pauli matrices acting on spin $j$: \textit{(i)} iterations from OBC, \textit{(ii)} iterations from  ``perfect-dephaser'' boundary condition (PDBC), i.e., $|\mathscr{I}_{0,T}\rangle=\bigotimes_{\tau=1}^T \frac 1 2 |\mathbb{1}\rangle
\otimes |\mathbb{1}\rangle
$ (cf. Refs.~\cite{lerose2020influence,sonner2021aop}), \textit{(iii)} LCGA. For each of them, we computed $S_{L,T}$ as a function of $L$ viewed as the iteration counter.
In agreement with previous work~\cite{sonner2021aop,Chan21}, we find a large TEB effect for OBC:
Approaching the peak, data become inaccurate (dashed lines), as the bond dimension cutoff (here $\chi=128$) is not sufficient to capture the massively entangled intermediate IMs. 
The PDBC partially mitigates this effect, as
it represents the IM of a maximally chaotic circuit~\cite{lerose2020influence}, closer to the
bulk IM of the non-integrable  model under consideration compared to the OBC IM {\it (i)}.
However, this $|\mathscr{I}_{0,T}\rangle$ also generates boundary ``reflections'' of quantum information, 
giving
rise to an extensive TEB~\footnote{The PDBC can be thought as generated by a dual-unitary circuit extending past the end of chain, at sites $j=L+1,L+2,\dots$~\cite{lerose2020influence}. The backflow of quantum information can then be thought as resulting from the {\it inhomogeneity} of the environment at $j=L$. For example, with refence to the quasiparticle picture outlined below, a quasiparticle reaching $j=L$ from $\mathcal{S}$ gets partially reflected back to $\mathcal{S}$ (and partially transmitted past $j=L$), which results in an extensive TEB. Due to partial transmission, in this case the peak is lower compared to the OBC case. 
}.
Crucially, in agreement with the above proof, TEB is fully circumvented by LCGA.

Furthermore, in Fig.~\ref{fig_benchmark} we compare the results of long-time simulation of local dynamics using LCGA and conventional time-evolving block decimation (TEBD)~\cite{vidaltebd}, for the same KIC prepared in a fully polarized state along $x$ direction (a), and at infinite temperature (b). The standard TEBD routine is run for the time-evolving wave function or for the Heisenberg-picture local operator, respectively.
In both cases, TEBD data deviate from the converged result when the entanglement wall is encountered (not shown), whereas LCGA reliably gives access to  long-time dynamics (at least in this model).

\textit{\bf Origin of temporal entanglement barrier.} 
%
\begin{figure}
    \centering
    \includegraphics[width=0.4\textwidth]{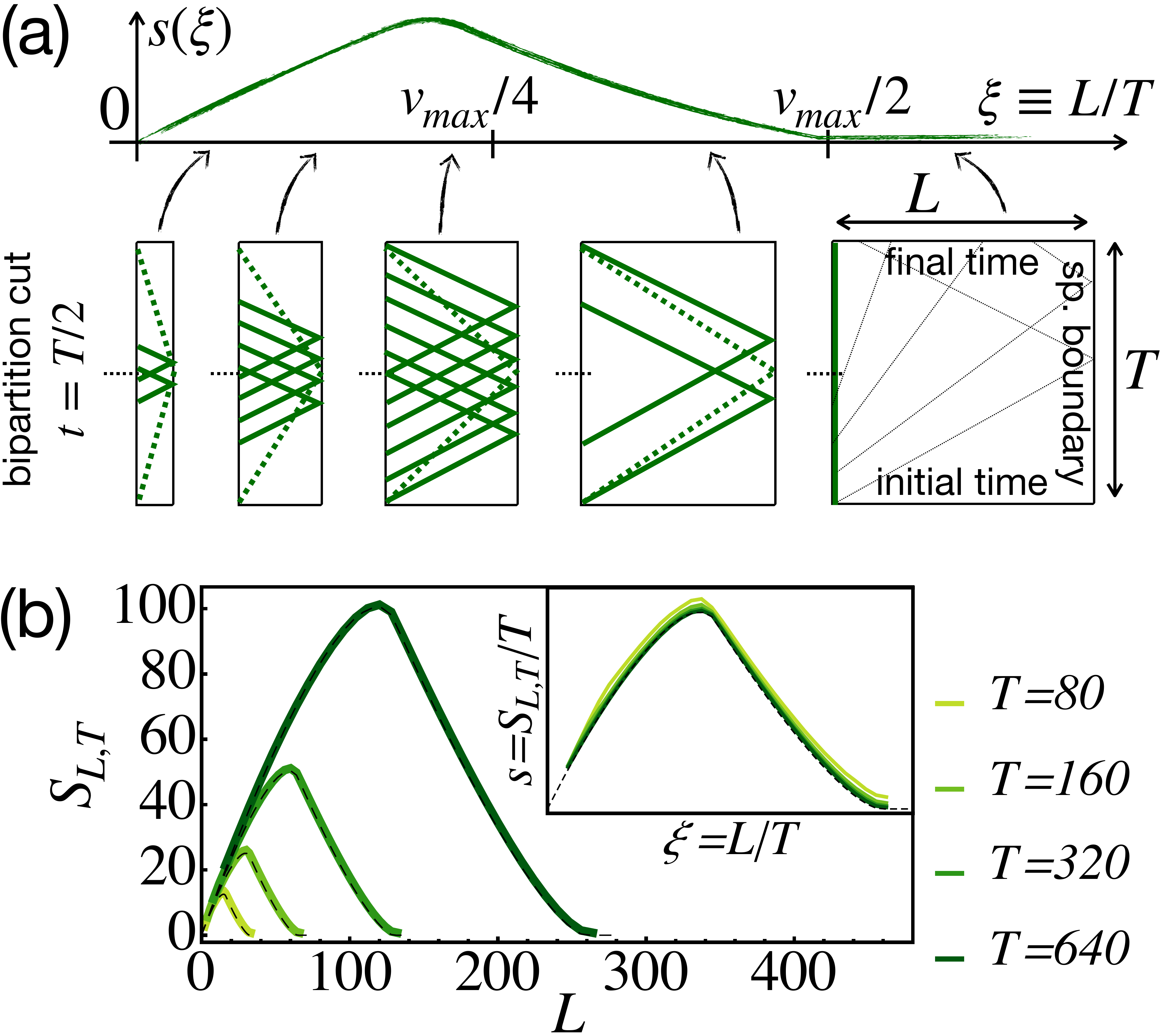}
    \caption{
        \textit{(a)}: Quasiparticle picture of TE scaling, using standard graphical convention of Ref.~\cite{Calabrese_2005}. Bottom: for IMs of right-environment with increasing size from left to right, 
        dashed (solid) lines denote the trajectories of slowest (fastest) quasiparticles creating correlations across the cut. 
        Top:  resulting characteristic TEB shape.
         {\it (b)}:
        Exact TEB of the OBC, infinite-temperature KIC for $g=\pi/4$, $J=0.6 \pi/4$. Dashed curves are the predictions of Eq.~\eqref{eq_qpformula} with $w \equiv 0.93 \log 2$ (essentially superimposed).
        The left asymptote is a straight line $\lim_{T\to\infty} S_{L,T}=v_{TE} L$.
        The right asymptote $
        S_\infty
        =
        \lim_{T\to\infty}
        \lim_{L\to\infty} S_{L,T}
        $ is a small finite quantity~\cite{lerose2021integrable}.
    }
    \label{fig_qp}
\end{figure}
%
To elucidate the physical origin of TEB, we consider a model with ballistic quasiparticles with maximum velocity $v_{\text{max}}$.
%
%
The dual ``evolution" $|\mathscr{I}_{\ell,T}\rangle$ vs $\ell$ described above, with space and time interchanged, is particularly transparent when the dual circuit obtained by a $90^\circ$ rotation of the original circuit, is unitary (see Ref.~\cite{Bertini2019}).
In this case, all quasiparticles move at the ``speed of light'' $v_{\text{max}}=2$, and one can adapt the Calabrese-Cardy picture of entanglement growth after a quantum quench~\cite{Calabrese_2005}. For OBC, $|\mathscr{I}_{0,T}\rangle$ is a product state of $2T$ Bell pairs [see Fig.~\ref{fig_IM}(a)], which is viewed as a source of pairs of entangled quasiparticles spreading with opposite velocities.
The bipartite entanglement of the final IM wavefunction $|\mathscr{I}_{L,T}\rangle$ between intervals $[0,t]$ and $[t,T]$ can be computed quasiclassically by summing the contributions of pairs propagating to points across the cut.
However, a novel aspect of our problem is given by the boundary conditions at $\tau=0$, $T$.
Their role is clearest in the original time-evolution perspective: a quasiparticle traveling with velocity $v$ on the forward sheet until $\tau=T$ continues along a superimposed, specular path on the backward sheet
(cf. the ``folding'' arguments in Refs.~\cite{Banuls09,muller2012tensor,giudice2021temporal}).
Hence, returning to the dual evolution picture, each point of the $\tau=T$ boundary can be viewed as 
{a source ($v>0$) and sink ($v<0$)} of superimposed quasiparticle pairs propagating on the two sheets.
For infinite-temperature initial state $\rho_{\mathcal{E}}(0)=\bigotimes_{j=1}^L \frac{\mathbb{1}}{2}$, the $\tau=0$ boundary plays the same role.
Thus, these boundaries destroy the non-local correlations of the pairs generated by $|\mathscr{I}_{0,T}\rangle$, replacing them with local correlations between the forward and backward branch at equal time.
As a result, when a quasiparticle hits a boundary (``absorption''), the corresponding initially entangled pair no longer contributes to TE, and has to be removed from the Calabrese-Cardy summation. 

The generation of a counter-propagating entangled pair in the quasiparticle picture appears as {\it reflection} of an individual quasiparticle in the dual picture.
Within the spatial evolution of $|\mathscr{I}\rangle$, pairs generated by a pure initial state $\rho_{\mathcal{E}}(0)=|\Psi_{\mathcal{E}}(0)\rangle\langle\Psi_{\mathcal{E}}(0)|$ at $\tau=0$~\cite{Calabrese_2005} are interpreted as reflections of individual quasiparticles hitting the boundary~\footnote{with a reflection coefficient related to the cross section of production}.
Analogously, within the physical time evolution, pairs generated by the spatial boundary $|\mathscr{I}_{0,T}\rangle$ are interpreted as {reflections} of individual quasiparticles hitting the end of the chain and coming back to the subsystem $\mathcal{S}$.
This elementary process leads to strong correlations between  the remote past and future of $\mathcal{S}$, and hence to a sizeable TEB.


By analogy with the conventional real-space entanglement spreading, the above picture is expected to hold for general models with stable quasiparticles, 
as illustrated in Fig.~\ref{fig_qp}(a).
We thus posit the following general formula for the  leading behavior of TE entropy
as $L\to\infty$, $T\to\infty$:
\begin{widetext}
\begin{equation}
\label{eq_qpformula}
    S_{L,T,t}=
    \int_0^t
    dt_{f,1}
    \int_t^T
    dt_{f,2}
    \int_0^T dt_{i}
     \int_{\omega_{\text{min}}}^{\omega_{\text{max}}} \frac{d\omega}{2\pi}
      \, w(\omega) \,
     \delta\bigg( t_{f,1}-t_i + \frac{L}{v(\omega)}
     \bigg)
     \delta\bigg( t_{f,2}-t_i - \frac{L}{v(\omega)}
     \bigg)
     \, .
\end{equation}
\end{widetext}
Here, $t_i$ denotes the boundary point where a quasiparticle pair is created, $t_{f,1}$, $t_{f,2}$ are the arrival points, $[\omega_{\text{min}},\omega_{\text{max}}]$ is the quasiparticle energy band, $v=d\omega/dk$ is the quasiparticle group velocity,
and $w$ accounts for the  entropy contribution
of a pair.

Setting $t=T/2$ for simplicity, Eq.~(\ref{eq_qpformula}) implies that $S_{L,T}/T \equiv s(\xi)$ is a function of the geometric parameter
$\xi=L/T$ only.
Taking the limit $T\to\infty$ first, one obtains a strictly linear increase $S=v_{TE} L$, with $v_{TE}=2\int_{\omega_{\text{min}}}^{\omega_{\text{max}}} \frac{d\omega}{2\pi } \frac{w(\omega)}{v(\omega)}$
\footnote{Note that $v(\omega)\sim \sqrt{\omega-\omega_{\text{min}}}$ or $\sqrt{\omega_{\text{max}}-\omega}$ at the quasienergy band edges, and the entropy contribution of pairs is bounded as $w(\omega)\le 2\log 2$ (the factor $2$ accounts for the two branches of the Keldysh contour). Thus, the integral that defines $v_{TE}$ is convergent.}.
However, for finite $\xi$, the boundaries absorb slow quasiparticles with $v<4\xi$, making the curve $s(\xi)$ bend down from the straight line $v_{TE} \xi$.
When $\xi$ exceeds $v_{\text{max}}/4$, quasiparticles at all velocities are being absorbed, so $s$ is now {\it decreasing} with $\xi$.
Finally, approaching
$v_{\text{max}}/2$,
only a vanishing fraction of fast quasiparticles avoid the boundaries. Thus, $s(\xi)$ has a barrier shape, sketched in Fig.~\ref{fig_qp}(a).
%
The infinite-system [$L>(v_{\text{max}}/2) T$
] TE entropy
$S_{\infty,T}$ can be  shown to saturate  as
$T\to\infty$ ({\it area-law}) in models with free ballistic quasiparticles~\cite{lerose2021integrable}.
The semiclassical formula~\eqref{eq_qpformula}
reproduces the extensive (\textit{volume-law}) term of the asymptotic scaling $S
\; {\sim} \; s(\xi) \cdot T + S_\infty $ as $T\to\infty$, which vanishes for $\xi\ge v_{\text{max}}/2$.


To verify this TEB scenario and to test Eq.~\eqref{eq_qpformula}, we computed $S_{L,T}$ in the transverse-field KIC [Eq.~\eqref{eq_KIC} with $h=0$],  
which can be mapped onto a non-interacting fermionic chain.
Figure~\ref{fig_qp}{(b)} provides a comparison between the exact result, 
obtained using Grassmann path integral
techniques~\cite{lerose2021integrable}, and the quasiparticle formula~\eqref{eq_qpformula}
with $w\equiv \text{const}$.
The weight $w\equiv \log 2$ makes Eq.~\eqref{eq_qpformula}
exact at the
self-dual points $|J|=|g|=\pi/4$ of the model~\cite{Guhr1}, where $s(\xi)= \log 2 \; {\rm min} (\xi,1/2-\xi)$.
The ansatz $w\equiv \text{const}$  captures  the properties of TEB -- including the location of its maximum -- fairly well throughout the phase diagram [see Fig.~\ref{fig_qp}(b)]
\footnote{
We note that the exact entropy function $w(\omega)$ cannot be expressed in terms of the microscopic data of the model by straightforwardly generalizing the standard arguments valid for unitary dynamics~\cite{AlbaPNAS}: the non-unitary generator $\hat {\mathcal{T}}$ is non-diagonalizable~\cite{lerose2020influence}, which makes the formulation of a generalized Gibbs ensemble for the dual dynamics {\it a priori} unclear.
It could presumably be obtained by generalizing the calculation of Ref.~\cite{CalabreseFagotti08} to non-unitary evolution.
}.

We note that the simplifying assumptions of an infinite-temperature initial state and OBC are not crucial for the TEB phenomenology: As is clear from the quasiparticle picture, neither the possible presence of counter-propagating entangled pairs at $\tau=0$~\cite{Calabrese_2005}, nor modified reflections at the edge of $\mathcal{E}$, affect the qualitative behavior
\footnote{Here we are  considering initial states with a pair structure of quasiparticles, as usually arising from global quenches~\cite{Calabrese_2005}.
We note that fine-tuned examples are known for which this is not the case~\cite{bertini2018entanglement,bastianello2018spreading}. In this case, a different behavior could be envisaged.
}.
We have explicitly checked this for different $\rho_{\mathcal{E}}$ and $|\mathscr{I}_{0,T}\rangle$ in the transverse-field KIC via exact
computations and numerical simulations [cf. Ref.~\cite{giudice2021temporal} and Fig.~\ref{fig_lc}(c)].

Finally, we have demonstrated that TEB persists when integrability is broken, and quasiparticles acquire a finite lifetime [Fig.~\ref{fig_lc}(c)], or may not even be well-defined. This shows that TEB  generally arises from back-flow of quantum information from finite environments, involving not only simple quasiparticle reflections from the boundary but also multi-quasiparticle scattering processes.

{\bf{Outlook.}}
The efficiency of the ``transverse contraction and folding'' MPS algorithm was  empirically observed in early applications (see Refs.~\cite{Banuls09,muller2012tensor} and~\cite{schollwoeckrev}).
Our results lay a firm physical basis for the origin of this efficiency in terms of the TEB picture, at the same time providing a more direct way to exploit it through the LCGA.
Thus, the the problem of inventing a truncation scheme for time evolution that preserves accuracy for long times~\cite{leviatan2017quantum,kloss2018tdvp,white2018dmt,surace2019simulating,ye2020emergenthydro,krumnow2019towards,rakovszky2020dissipation,Karrasch_2013,hauschild2018finding} can be reduced to a well-defined problem: understanding under which conditions a quantum many-body system can be faithfully represented as an effective ``small'' bath.
Understanding the scaling the optimal bath dimension $\chi$ with time for generic chaotic quantum many-body systems remains an outstanding challenge for future work. 

%

This work reveals a novel aspect of space-time duality~\cite{Guhr1,BertiniSFF,Bertini2019,ippoliti2021postselection,grover2021spacetime,ippoliti2021fractal,chan2021lyapunov,garratt2021pairing,garratt2021mbdl}, as the TEB effect is absent in
the pure-state spacetime-dual
evolution discussed in the context of measurement-induced entanglement transitions~\cite{ippoliti2021fractal,grover2021spacetime}.
Furthermore, exploration of quantum dissipative dynamics using the LCGA  also represents a promising future direction~\cite{mi2022noise}.
Finally, an important issue -- which will be addressed elsewhere -- is  the general relation between TEB and real-space entanglement barriers~\cite{Dubail_2017,wang2019barrier,reid2021barrier}.

\begin{acknowledgments}
{\bf Acknowledgments.}
This work was supported by the Swiss National Science Foundation (SNSF) and by the European Research Council (ERC) under the European Union's Horizon 2020 research and innovation programme (grant agreement No. 864597).
\end{acknowledgments}

{\bf Note added.}
Reference~\cite{frias2022light}, which appeared shortly after this work, develops an efficient version of the LCGA for Hamiltonian dynamics, where the unphysical light velocity of the Trotterized circuit $v_{\text{lc}}=2/\Delta t \gg 1$ is replaced by the physical Lieb-Robinson velocity $v_{\text{max}} = \mathcal{O}(1)$.

\bibliography{mbl}

\begin{thebibliography}{52}%
\makeatletter
\providecommand \@ifxundefined [1]{%
 \@ifx{#1\undefined}
}%
\providecommand \@ifnum [1]{%
 \ifnum #1\expandafter \@firstoftwo
 \else \expandafter \@secondoftwo
 \fi
}%
\providecommand \@ifx [1]{%
 \ifx #1\expandafter \@firstoftwo
 \else \expandafter \@secondoftwo
 \fi
}%
\providecommand \natexlab [1]{#1}%
\providecommand \enquote  [1]{``#1''}%
\providecommand \bibnamefont  [1]{#1}%
\providecommand \bibfnamefont [1]{#1}%
\providecommand \citenamefont [1]{#1}%
\providecommand \href@noop [0]{\@secondoftwo}%
\providecommand \href [0]{\begingroup \@sanitize@url \@href}%
\providecommand \@href[1]{\@@startlink{#1}\@@href}%
\providecommand \@@href[1]{\endgroup#1\@@endlink}%
\providecommand \@sanitize@url [0]{\catcode `\\12\catcode `\$12\catcode
  `\&12\catcode `\#12\catcode `\^12\catcode `\_12\catcode `\%12\relax}%
\providecommand \@@startlink[1]{}%
\providecommand \@@endlink[0]{}%
\providecommand \url  [0]{\begingroup\@sanitize@url \@url }%
\providecommand \@url [1]{\endgroup\@href {#1}{\urlprefix }}%
\providecommand \urlprefix  [0]{URL }%
\providecommand \Eprint [0]{\href }%
\providecommand \doibase [0]{http://dx.doi.org/}%
\providecommand \selectlanguage [0]{\@gobble}%
\providecommand \bibinfo  [0]{\@secondoftwo}%
\providecommand \bibfield  [0]{\@secondoftwo}%
\providecommand \translation [1]{[#1]}%
\providecommand \BibitemOpen [0]{}%
\providecommand \bibitemStop [0]{}%
\providecommand \bibitemNoStop [0]{.\EOS\space}%
\providecommand \EOS [0]{\spacefactor3000\relax}%
\providecommand \BibitemShut  [1]{\csname bibitem#1\endcsname}%
\let\auto@bib@innerbib\@empty
\bibitem [{\citenamefont {Schollw{\"o}ck}(2011)}]{schollwoeckrev}%
  \BibitemOpen
  \bibfield  {author} {\bibinfo {author} {\bibfnamefont {U.}~\bibnamefont
  {Schollw{\"o}ck}},\ }\href {\doibase
  https://doi.org/10.1016/j.aop.2010.09.012} {\bibfield  {journal} {\bibinfo
  {journal} {Annals of physics}\ }\textbf {\bibinfo {volume} {326}},\ \bibinfo
  {pages} {96} (\bibinfo {year} {2011})}\BibitemShut {NoStop}%
\bibitem [{\citenamefont {Calabrese}\ and\ \citenamefont
  {Cardy}(2005)}]{Calabrese_2005}%
  \BibitemOpen
  \bibfield  {author} {\bibinfo {author} {\bibfnamefont {P.}~\bibnamefont
  {Calabrese}}\ and\ \bibinfo {author} {\bibfnamefont {J.}~\bibnamefont
  {Cardy}},\ }\href {\doibase 10.1088/1742-5468/2005/04/p04010} {\bibfield
  {journal} {\bibinfo  {journal} {Journal of Statistical Mechanics: Theory and
  Experiment}\ }\textbf {\bibinfo {volume} {2005}},\ \bibinfo {pages} {P04010}
  (\bibinfo {year} {2005})}\BibitemShut {NoStop}%
\bibitem [{\citenamefont {Kim}\ and\ \citenamefont {Huse}(2013)}]{Kim13}%
  \BibitemOpen
  \bibfield  {author} {\bibinfo {author} {\bibfnamefont {H.}~\bibnamefont
  {Kim}}\ and\ \bibinfo {author} {\bibfnamefont {D.~A.}\ \bibnamefont {Huse}},\
  }\href {\doibase 10.1103/PhysRevLett.111.127205} {\bibfield  {journal}
  {\bibinfo  {journal} {Phys. Rev. Lett.}\ }\textbf {\bibinfo {volume} {111}},\
  \bibinfo {pages} {127205} (\bibinfo {year} {2013})}\BibitemShut {NoStop}%
\bibitem [{\citenamefont {Nahum}\ \emph {et~al.}(2017)\citenamefont {Nahum},
  \citenamefont {Ruhman}, \citenamefont {Vijay},\ and\ \citenamefont
  {Haah}}]{NahumPRXEntanglement}%
  \BibitemOpen
  \bibfield  {author} {\bibinfo {author} {\bibfnamefont {A.}~\bibnamefont
  {Nahum}}, \bibinfo {author} {\bibfnamefont {J.}~\bibnamefont {Ruhman}},
  \bibinfo {author} {\bibfnamefont {S.}~\bibnamefont {Vijay}}, \ and\ \bibinfo
  {author} {\bibfnamefont {J.}~\bibnamefont {Haah}},\ }\href {\doibase
  10.1103/PhysRevX.7.031016} {\bibfield  {journal} {\bibinfo  {journal} {Phys.
  Rev. X}\ }\textbf {\bibinfo {volume} {7}},\ \bibinfo {pages} {031016}
  (\bibinfo {year} {2017})}\BibitemShut {NoStop}%
\bibitem [{\citenamefont {Leviatan}\ \emph {et~al.}(2017)\citenamefont
  {Leviatan}, \citenamefont {Pollmann}, \citenamefont {Bardarson},
  \citenamefont {Huse},\ and\ \citenamefont {Altman}}]{leviatan2017quantum}%
  \BibitemOpen
  \bibfield  {author} {\bibinfo {author} {\bibfnamefont {E.}~\bibnamefont
  {Leviatan}}, \bibinfo {author} {\bibfnamefont {F.}~\bibnamefont {Pollmann}},
  \bibinfo {author} {\bibfnamefont {J.~H.}\ \bibnamefont {Bardarson}}, \bibinfo
  {author} {\bibfnamefont {D.~A.}\ \bibnamefont {Huse}}, \ and\ \bibinfo
  {author} {\bibfnamefont {E.}~\bibnamefont {Altman}},\ }\href
  {https://arxiv.org/abs/1702.08894} {\bibfield  {journal} {\bibinfo  {journal}
  {arXiv preprint arXiv:1702.08894}\ } (\bibinfo {year} {2017})}\BibitemShut
  {NoStop}%
\bibitem [{\citenamefont {Kloss}\ \emph {et~al.}(2018)\citenamefont {Kloss},
  \citenamefont {Lev},\ and\ \citenamefont {Reichman}}]{kloss2018tdvp}%
  \BibitemOpen
  \bibfield  {author} {\bibinfo {author} {\bibfnamefont {B.}~\bibnamefont
  {Kloss}}, \bibinfo {author} {\bibfnamefont {Y.~B.}\ \bibnamefont {Lev}}, \
  and\ \bibinfo {author} {\bibfnamefont {D.}~\bibnamefont {Reichman}},\ }\href
  {\doibase 10.1103/PhysRevB.97.024307} {\bibfield  {journal} {\bibinfo
  {journal} {Phys. Rev. B}\ }\textbf {\bibinfo {volume} {97}},\ \bibinfo
  {pages} {024307} (\bibinfo {year} {2018})}\BibitemShut {NoStop}%
\bibitem [{\citenamefont {White}\ \emph {et~al.}(2018)\citenamefont {White},
  \citenamefont {Zaletel}, \citenamefont {Mong},\ and\ \citenamefont
  {Refael}}]{white2018dmt}%
  \BibitemOpen
  \bibfield  {author} {\bibinfo {author} {\bibfnamefont {C.~D.}\ \bibnamefont
  {White}}, \bibinfo {author} {\bibfnamefont {M.}~\bibnamefont {Zaletel}},
  \bibinfo {author} {\bibfnamefont {R.~S.~K.}\ \bibnamefont {Mong}}, \ and\
  \bibinfo {author} {\bibfnamefont {G.}~\bibnamefont {Refael}},\ }\href
  {\doibase 10.1103/PhysRevB.97.035127} {\bibfield  {journal} {\bibinfo
  {journal} {Phys. Rev. B}\ }\textbf {\bibinfo {volume} {97}},\ \bibinfo
  {pages} {035127} (\bibinfo {year} {2018})}\BibitemShut {NoStop}%
\bibitem [{\citenamefont {Surace}\ \emph {et~al.}(2019)\citenamefont {Surace},
  \citenamefont {Piani},\ and\ \citenamefont
  {Tagliacozzo}}]{surace2019simulating}%
  \BibitemOpen
  \bibfield  {author} {\bibinfo {author} {\bibfnamefont {J.}~\bibnamefont
  {Surace}}, \bibinfo {author} {\bibfnamefont {M.}~\bibnamefont {Piani}}, \
  and\ \bibinfo {author} {\bibfnamefont {L.}~\bibnamefont {Tagliacozzo}},\
  }\href {\doibase 10.1103/PhysRevB.99.235115} {\bibfield  {journal} {\bibinfo
  {journal} {Phys. Rev. B}\ }\textbf {\bibinfo {volume} {99}},\ \bibinfo
  {pages} {235115} (\bibinfo {year} {2019})}\BibitemShut {NoStop}%
\bibitem [{\citenamefont {Ye}\ \emph {et~al.}(2020)\citenamefont {Ye},
  \citenamefont {Machado}, \citenamefont {White}, \citenamefont {Mong},\ and\
  \citenamefont {Yao}}]{ye2020emergenthydro}%
  \BibitemOpen
  \bibfield  {author} {\bibinfo {author} {\bibfnamefont {B.}~\bibnamefont
  {Ye}}, \bibinfo {author} {\bibfnamefont {F.}~\bibnamefont {Machado}},
  \bibinfo {author} {\bibfnamefont {C.~D.}\ \bibnamefont {White}}, \bibinfo
  {author} {\bibfnamefont {R.~S.~K.}\ \bibnamefont {Mong}}, \ and\ \bibinfo
  {author} {\bibfnamefont {N.~Y.}\ \bibnamefont {Yao}},\ }\href {\doibase
  10.1103/PhysRevLett.125.030601} {\bibfield  {journal} {\bibinfo  {journal}
  {Phys. Rev. Lett.}\ }\textbf {\bibinfo {volume} {125}},\ \bibinfo {pages}
  {030601} (\bibinfo {year} {2020})}\BibitemShut {NoStop}%
\bibitem [{\citenamefont {Krumnow}\ \emph {et~al.}(2019)\citenamefont
  {Krumnow}, \citenamefont {Eisert},\ and\ \citenamefont
  {Legeza}}]{krumnow2019towards}%
  \BibitemOpen
  \bibfield  {author} {\bibinfo {author} {\bibfnamefont {C.}~\bibnamefont
  {Krumnow}}, \bibinfo {author} {\bibfnamefont {J.}~\bibnamefont {Eisert}}, \
  and\ \bibinfo {author} {\bibfnamefont {{\"O}.}~\bibnamefont {Legeza}},\
  }\href {https://arxiv.org/abs/1904.11999} {\bibfield  {journal} {\bibinfo
  {journal} {arXiv preprint arXiv:1904.11999}\ } (\bibinfo {year}
  {2019})}\BibitemShut {NoStop}%
\bibitem [{\citenamefont {Rakovszky}\ \emph {et~al.}(2020)\citenamefont
  {Rakovszky}, \citenamefont {von Keyserlingk},\ and\ \citenamefont
  {Pollmann}}]{rakovszky2020dissipation}%
  \BibitemOpen
  \bibfield  {author} {\bibinfo {author} {\bibfnamefont {T.}~\bibnamefont
  {Rakovszky}}, \bibinfo {author} {\bibfnamefont {C.}~\bibnamefont {von
  Keyserlingk}}, \ and\ \bibinfo {author} {\bibfnamefont {F.}~\bibnamefont
  {Pollmann}},\ }\href {https://arxiv.org/abs/2004.05177} {\bibfield  {journal}
  {\bibinfo  {journal} {arXiv preprint arXiv:2004.05177}\ } (\bibinfo {year}
  {2020})}\BibitemShut {NoStop}%
\bibitem [{\citenamefont {Karrasch}\ \emph {et~al.}(2013)\citenamefont
  {Karrasch}, \citenamefont {Bardarson},\ and\ \citenamefont
  {Moore}}]{Karrasch_2013}%
  \BibitemOpen
  \bibfield  {author} {\bibinfo {author} {\bibfnamefont {C.}~\bibnamefont
  {Karrasch}}, \bibinfo {author} {\bibfnamefont {J.~H.}\ \bibnamefont
  {Bardarson}}, \ and\ \bibinfo {author} {\bibfnamefont {J.~E.}\ \bibnamefont
  {Moore}},\ }\href {\doibase 10.1088/1367-2630/15/8/083031} {\bibfield
  {journal} {\bibinfo  {journal} {New Journal of Physics}\ }\textbf {\bibinfo
  {volume} {15}},\ \bibinfo {pages} {083031} (\bibinfo {year}
  {2013})}\BibitemShut {NoStop}%
\bibitem [{\citenamefont {Hauschild}\ \emph {et~al.}(2018)\citenamefont
  {Hauschild}, \citenamefont {Leviatan}, \citenamefont {Bardarson},
  \citenamefont {Altman}, \citenamefont {Zaletel},\ and\ \citenamefont
  {Pollmann}}]{hauschild2018finding}%
  \BibitemOpen
  \bibfield  {author} {\bibinfo {author} {\bibfnamefont {J.}~\bibnamefont
  {Hauschild}}, \bibinfo {author} {\bibfnamefont {E.}~\bibnamefont {Leviatan}},
  \bibinfo {author} {\bibfnamefont {J.~H.}\ \bibnamefont {Bardarson}}, \bibinfo
  {author} {\bibfnamefont {E.}~\bibnamefont {Altman}}, \bibinfo {author}
  {\bibfnamefont {M.~P.}\ \bibnamefont {Zaletel}}, \ and\ \bibinfo {author}
  {\bibfnamefont {F.}~\bibnamefont {Pollmann}},\ }\href {\doibase
  10.1103/PhysRevB.98.235163} {\bibfield  {journal} {\bibinfo  {journal} {Phys.
  Rev. B}\ }\textbf {\bibinfo {volume} {98}},\ \bibinfo {pages} {235163}
  (\bibinfo {year} {2018})}\BibitemShut {NoStop}%
\bibitem [{\citenamefont {Feynman}\ and\ \citenamefont
  {Vernon}(1963)}]{FeynmanVernon}%
  \BibitemOpen
  \bibfield  {author} {\bibinfo {author} {\bibfnamefont {R.}~\bibnamefont
  {Feynman}}\ and\ \bibinfo {author} {\bibfnamefont {F.}~\bibnamefont
  {Vernon}},\ }\href {\doibase https://doi.org/10.1016/0003-4916(63)90068-X}
  {\bibfield  {journal} {\bibinfo  {journal} {Annals of Physics}\ }\textbf
  {\bibinfo {volume} {24}},\ \bibinfo {pages} {118 } (\bibinfo {year}
  {1963})}\BibitemShut {NoStop}%
\bibitem [{\citenamefont {Leggett}\ \emph {et~al.}(1987)\citenamefont
  {Leggett}, \citenamefont {Chakravarty}, \citenamefont {Dorsey}, \citenamefont
  {Fisher}, \citenamefont {Garg},\ and\ \citenamefont {Zwerger}}]{LeggettRMP}%
  \BibitemOpen
  \bibfield  {author} {\bibinfo {author} {\bibfnamefont {A.~J.}\ \bibnamefont
  {Leggett}}, \bibinfo {author} {\bibfnamefont {S.}~\bibnamefont
  {Chakravarty}}, \bibinfo {author} {\bibfnamefont {A.~T.}\ \bibnamefont
  {Dorsey}}, \bibinfo {author} {\bibfnamefont {M.~P.~A.}\ \bibnamefont
  {Fisher}}, \bibinfo {author} {\bibfnamefont {A.}~\bibnamefont {Garg}}, \ and\
  \bibinfo {author} {\bibfnamefont {W.}~\bibnamefont {Zwerger}},\ }\href
  {\doibase 10.1103/RevModPhys.59.1} {\bibfield  {journal} {\bibinfo  {journal}
  {Rev. Mod. Phys.}\ }\textbf {\bibinfo {volume} {59}},\ \bibinfo {pages} {1}
  (\bibinfo {year} {1987})}\BibitemShut {NoStop}%
\bibitem [{\citenamefont {Lerose}\ \emph
  {et~al.}(2021{\natexlab{a}})\citenamefont {Lerose}, \citenamefont {Sonner},\
  and\ \citenamefont {Abanin}}]{lerose2020influence}%
  \BibitemOpen
  \bibfield  {author} {\bibinfo {author} {\bibfnamefont {A.}~\bibnamefont
  {Lerose}}, \bibinfo {author} {\bibfnamefont {M.}~\bibnamefont {Sonner}}, \
  and\ \bibinfo {author} {\bibfnamefont {D.~A.}\ \bibnamefont {Abanin}},\
  }\href {\doibase 10.1103/PhysRevX.11.021040} {\bibfield  {journal} {\bibinfo
  {journal} {Phys. Rev. X}\ }\textbf {\bibinfo {volume} {11}},\ \bibinfo
  {pages} {021040} (\bibinfo {year} {2021}{\natexlab{a}})}\BibitemShut
  {NoStop}%
\bibitem [{\citenamefont {Sonner}\ \emph {et~al.}(2021)\citenamefont {Sonner},
  \citenamefont {Lerose},\ and\ \citenamefont {Abanin}}]{sonner2021aop}%
  \BibitemOpen
  \bibfield  {author} {\bibinfo {author} {\bibfnamefont {M.}~\bibnamefont
  {Sonner}}, \bibinfo {author} {\bibfnamefont {A.}~\bibnamefont {Lerose}}, \
  and\ \bibinfo {author} {\bibfnamefont {D.~A.}\ \bibnamefont {Abanin}},\
  }\href {\doibase https://doi.org/10.1016/j.aop.2021.168677} {\bibfield
  {journal} {\bibinfo  {journal} {Annals of Physics}\ }\textbf {\bibinfo
  {volume} {435}},\ \bibinfo {pages} {168677} (\bibinfo {year} {2021})},\
  \bibinfo {note} {special issue on Philip W. Anderson}\BibitemShut {NoStop}%
\bibitem [{\citenamefont {Ba\~nuls}\ \emph {et~al.}(2009)\citenamefont
  {Ba\~nuls}, \citenamefont {Hastings}, \citenamefont {Verstraete},\ and\
  \citenamefont {Cirac}}]{Banuls09}%
  \BibitemOpen
  \bibfield  {author} {\bibinfo {author} {\bibfnamefont {M.~C.}\ \bibnamefont
  {Ba\~nuls}}, \bibinfo {author} {\bibfnamefont {M.~B.}\ \bibnamefont
  {Hastings}}, \bibinfo {author} {\bibfnamefont {F.}~\bibnamefont
  {Verstraete}}, \ and\ \bibinfo {author} {\bibfnamefont {J.~I.}\ \bibnamefont
  {Cirac}},\ }\href {\doibase 10.1103/PhysRevLett.102.240603} {\bibfield
  {journal} {\bibinfo  {journal} {Phys. Rev. Lett.}\ }\textbf {\bibinfo
  {volume} {102}},\ \bibinfo {pages} {240603} (\bibinfo {year}
  {2009})}\BibitemShut {NoStop}%
\bibitem [{\citenamefont {Lerose}\ \emph
  {et~al.}(2021{\natexlab{b}})\citenamefont {Lerose}, \citenamefont {Sonner},\
  and\ \citenamefont {Abanin}}]{lerose2021integrable}%
  \BibitemOpen
  \bibfield  {author} {\bibinfo {author} {\bibfnamefont {A.}~\bibnamefont
  {Lerose}}, \bibinfo {author} {\bibfnamefont {M.}~\bibnamefont {Sonner}}, \
  and\ \bibinfo {author} {\bibfnamefont {D.~A.}\ \bibnamefont {Abanin}},\
  }\href {\doibase 10.1103/PhysRevB.104.035137} {\bibfield  {journal} {\bibinfo
   {journal} {Phys. Rev. B}\ }\textbf {\bibinfo {volume} {104}},\ \bibinfo
  {pages} {035137} (\bibinfo {year} {2021}{\natexlab{b}})}\BibitemShut
  {NoStop}%
\bibitem [{\citenamefont {Piroli}\ \emph {et~al.}(2020)\citenamefont {Piroli},
  \citenamefont {Bertini}, \citenamefont {Cirac},\ and\ \citenamefont
  {Prosen}}]{Piroli2020exact}%
  \BibitemOpen
  \bibfield  {author} {\bibinfo {author} {\bibfnamefont {L.}~\bibnamefont
  {Piroli}}, \bibinfo {author} {\bibfnamefont {B.}~\bibnamefont {Bertini}},
  \bibinfo {author} {\bibfnamefont {J.~I.}\ \bibnamefont {Cirac}}, \ and\
  \bibinfo {author} {\bibfnamefont {T.~c.~v.}\ \bibnamefont {Prosen}},\ }\href
  {\doibase 10.1103/PhysRevB.101.094304} {\bibfield  {journal} {\bibinfo
  {journal} {Phys. Rev. B}\ }\textbf {\bibinfo {volume} {101}},\ \bibinfo
  {pages} {094304} (\bibinfo {year} {2020})}\BibitemShut {NoStop}%
\bibitem [{\citenamefont {Klobas}\ \emph {et~al.}(2021)\citenamefont {Klobas},
  \citenamefont {Bertini},\ and\ \citenamefont {Piroli}}]{Koblas20}%
  \BibitemOpen
  \bibfield  {author} {\bibinfo {author} {\bibfnamefont {K.}~\bibnamefont
  {Klobas}}, \bibinfo {author} {\bibfnamefont {B.}~\bibnamefont {Bertini}}, \
  and\ \bibinfo {author} {\bibfnamefont {L.}~\bibnamefont {Piroli}},\ }\href
  {\doibase 10.1103/PhysRevLett.126.160602} {\bibfield  {journal} {\bibinfo
  {journal} {Phys. Rev. Lett.}\ }\textbf {\bibinfo {volume} {126}},\ \bibinfo
  {pages} {160602} (\bibinfo {year} {2021})}\BibitemShut {NoStop}%
\bibitem [{\citenamefont {Klobas}\ and\ \citenamefont
  {Bertini}(2021)}]{Klobas2021exact}%
  \BibitemOpen
  \bibfield  {author} {\bibinfo {author} {\bibfnamefont {K.}~\bibnamefont
  {Klobas}}\ and\ \bibinfo {author} {\bibfnamefont {B.}~\bibnamefont
  {Bertini}},\ }\href {\doibase 10.21468/SciPostPhys.11.6.106} {\bibfield
  {journal} {\bibinfo  {journal} {SciPost Phys.}\ }\textbf {\bibinfo {volume}
  {11}},\ \bibinfo {pages} {106} (\bibinfo {year} {2021})}\BibitemShut
  {NoStop}%
\bibitem [{\citenamefont {Giudice}\ \emph {et~al.}(2021)\citenamefont
  {Giudice}, \citenamefont {Giudici}, \citenamefont {Sonner}, \citenamefont
  {Thoenniss}, \citenamefont {Lerose}, \citenamefont {Abanin},\ and\
  \citenamefont {Piroli}}]{giudice2021temporal}%
  \BibitemOpen
  \bibfield  {author} {\bibinfo {author} {\bibfnamefont {G.}~\bibnamefont
  {Giudice}}, \bibinfo {author} {\bibfnamefont {G.}~\bibnamefont {Giudici}},
  \bibinfo {author} {\bibfnamefont {M.}~\bibnamefont {Sonner}}, \bibinfo
  {author} {\bibfnamefont {J.}~\bibnamefont {Thoenniss}}, \bibinfo {author}
  {\bibfnamefont {A.}~\bibnamefont {Lerose}}, \bibinfo {author} {\bibfnamefont
  {D.~A.}\ \bibnamefont {Abanin}}, \ and\ \bibinfo {author} {\bibfnamefont
  {L.}~\bibnamefont {Piroli}},\ }\href {https://arxiv.org/abs/2112.14264}
  {\bibfield  {journal} {\bibinfo  {journal} {arXiv preprint arXiv:2112.14264}\
  } (\bibinfo {year} {2021})}\BibitemShut {NoStop}%
\bibitem [{\citenamefont {{Sonner}}\ \emph {et~al.}(2020)\citenamefont
  {{Sonner}}, \citenamefont {{Lerose}},\ and\ \citenamefont
  {{Abanin}}}]{Sonner20CharacterizingMBL}%
  \BibitemOpen
  \bibfield  {author} {\bibinfo {author} {\bibfnamefont {M.}~\bibnamefont
  {{Sonner}}}, \bibinfo {author} {\bibfnamefont {A.}~\bibnamefont {{Lerose}}},
  \ and\ \bibinfo {author} {\bibfnamefont {D.~A.}\ \bibnamefont {{Abanin}}},\
  }\href@noop {} {\bibfield  {journal} {\bibinfo  {journal} {arXiv e-prints}\
  ,\ \bibinfo {eid} {arXiv:2012.00777}} (\bibinfo {year} {2020})},\ \Eprint
  {http://arxiv.org/abs/2012.00777} {arXiv:2012.00777 [cond-mat.dis-nn]}
  \BibitemShut {NoStop}%
\bibitem [{\citenamefont {Ye}\ and\ \citenamefont {Chan}(2021)}]{Chan21}%
  \BibitemOpen
  \bibfield  {author} {\bibinfo {author} {\bibfnamefont {E.}~\bibnamefont
  {Ye}}\ and\ \bibinfo {author} {\bibfnamefont {G.~K.-L.}\ \bibnamefont
  {Chan}},\ }\href {\doibase 10.1063/5.0047260} {\bibfield  {journal} {\bibinfo
   {journal} {The Journal of Chemical Physics}\ }\textbf {\bibinfo {volume}
  {155}},\ \bibinfo {pages} {044104} (\bibinfo {year} {2021})},\ \Eprint
  {http://arxiv.org/abs/https://doi.org/10.1063/5.0047260}
  {https://doi.org/10.1063/5.0047260} \BibitemShut {NoStop}%
\bibitem [{\citenamefont {M{\"u}ller-Hermes}\ \emph {et~al.}(2012)\citenamefont
  {M{\"u}ller-Hermes}, \citenamefont {Cirac},\ and\ \citenamefont
  {Banuls}}]{muller2012tensor}%
  \BibitemOpen
  \bibfield  {author} {\bibinfo {author} {\bibfnamefont {A.}~\bibnamefont
  {M{\"u}ller-Hermes}}, \bibinfo {author} {\bibfnamefont {J.~I.}\ \bibnamefont
  {Cirac}}, \ and\ \bibinfo {author} {\bibfnamefont {M.~C.}\ \bibnamefont
  {Banuls}},\ }\href {\doibase 10.1088/1367-2630/14/7/075003} {\bibfield
  {journal} {\bibinfo  {journal} {New Journal of Physics}\ }\textbf {\bibinfo
  {volume} {14}},\ \bibinfo {pages} {075003} (\bibinfo {year}
  {2012})}\BibitemShut {NoStop}%
\bibitem [{\citenamefont {Vidal}(2004)}]{vidaltebd}%
  \BibitemOpen
  \bibfield  {author} {\bibinfo {author} {\bibfnamefont {G.}~\bibnamefont
  {Vidal}},\ }\href {\doibase 10.1103/PhysRevLett.93.040502} {\bibfield
  {journal} {\bibinfo  {journal} {Phys. Rev. Lett.}\ }\textbf {\bibinfo
  {volume} {93}},\ \bibinfo {pages} {040502} (\bibinfo {year}
  {2004})}\BibitemShut {NoStop}%
\bibitem [{\citenamefont {Lieb}\ and\ \citenamefont
  {Robinson}(1972)}]{LiebRobinson}%
  \BibitemOpen
  \bibfield  {author} {\bibinfo {author} {\bibfnamefont {E.~H.}\ \bibnamefont
  {Lieb}}\ and\ \bibinfo {author} {\bibfnamefont {D.}~\bibnamefont
  {Robinson}},\ }\href {\doibase https://doi.org/10.1007/BF01645779} {\bibfield
   {journal} {\bibinfo  {journal} {Commun. Math. Phys.}\ }\textbf {\bibinfo
  {volume} {28}},\ \bibinfo {pages} {251} (\bibinfo {year} {1972})}\BibitemShut
  {NoStop}%
\bibitem [{\citenamefont {{De Roeck}}\ and\ \citenamefont
  {{Verreet}}(2019)}]{DeRoeckVerreet19}%
  \BibitemOpen
  \bibfield  {author} {\bibinfo {author} {\bibfnamefont {W.}~\bibnamefont {{De
  Roeck}}}\ and\ \bibinfo {author} {\bibfnamefont {V.}~\bibnamefont
  {{Verreet}}},\ }\href@noop {} {\bibfield  {journal} {\bibinfo  {journal}
  {arXiv e-prints}\ ,\ \bibinfo {eid} {arXiv:1911.01998}} (\bibinfo {year}
  {2019})},\ \Eprint {http://arxiv.org/abs/1911.01998} {arXiv:1911.01998
  [cond-mat.stat-mech]} \BibitemShut {NoStop}%
\bibitem [{Note1()}]{Note1}%
  \BibitemOpen
  \bibinfo {note} {The PDBC can be thought as generated by a dual-unitary
  circuit extending past the end of chain, at sites $j=L+1,L+2,\protect \dots
  $~\cite {lerose2020influence}. The backflow of quantum information can then
  be thought as resulting from the {\protect \it inhomogeneity} of the
  environment at $j=L$. For example, with refence to the quasiparticle picture
  outlined below, a quasiparticle reaching $j=L$ from $\protect \mathcal {S}$
  gets partially reflected back to $\protect \mathcal {S}$ (and partially
  transmitted past $j=L$), which results in an extensive TEB. Due to partial
  transmission, in this case the peak is lower compared to the OBC
  case.}\BibitemShut {Stop}%
\bibitem [{\citenamefont {Bertini}\ \emph {et~al.}(2019)\citenamefont
  {Bertini}, \citenamefont {Kos},\ and\ \citenamefont {Prosen}}]{Bertini2019}%
  \BibitemOpen
  \bibfield  {author} {\bibinfo {author} {\bibfnamefont {B.}~\bibnamefont
  {Bertini}}, \bibinfo {author} {\bibfnamefont {P.}~\bibnamefont {Kos}}, \ and\
  \bibinfo {author} {\bibfnamefont {T.~c.~v.}\ \bibnamefont {Prosen}},\ }\href
  {\doibase 10.1103/PhysRevLett.123.210601} {\bibfield  {journal} {\bibinfo
  {journal} {Phys. Rev. Lett.}\ }\textbf {\bibinfo {volume} {123}},\ \bibinfo
  {pages} {210601} (\bibinfo {year} {2019})}\BibitemShut {NoStop}%
\bibitem [{Note2()}]{Note2}%
  \BibitemOpen
  \bibinfo {note} {With a reflection coefficient related to the cross section
  of production}\BibitemShut {NoStop}%
\bibitem [{Note3()}]{Note3}%
  \BibitemOpen
  \bibinfo {note} {Note that $v(\omega )\thicksim \protect \sqrt {\omega
  -\omega _{\protect \text {min}}}$ or $\protect \sqrt {\omega _{\protect \text
  {max}}-\omega }$ at the quasienergy band edges, and the entropy contribution
  of pairs is bounded as $w(\omega )\le 2\protect \qopname \relax o{log}2$ (the
  factor $2$ accounts for the two branches of the Keldysh contour). Thus, the
  integral that defines $v_{TE}$ is convergent.}\BibitemShut {Stop}%
\bibitem [{\citenamefont {Akila}\ \emph {et~al.}(2016)\citenamefont {Akila},
  \citenamefont {Waltner}, \citenamefont {Gutkin},\ and\ \citenamefont
  {Guhr}}]{Guhr1}%
  \BibitemOpen
  \bibfield  {author} {\bibinfo {author} {\bibfnamefont {M.}~\bibnamefont
  {Akila}}, \bibinfo {author} {\bibfnamefont {D.}~\bibnamefont {Waltner}},
  \bibinfo {author} {\bibfnamefont {B.}~\bibnamefont {Gutkin}}, \ and\ \bibinfo
  {author} {\bibfnamefont {T.}~\bibnamefont {Guhr}},\ }\href {\doibase
  10.1088/1751-8113/49/37/375101} {\bibfield  {journal} {\bibinfo  {journal}
  {Journal of Physics A: Mathematical and Theoretical}\ }\textbf {\bibinfo
  {volume} {49}},\ \bibinfo {pages} {375101} (\bibinfo {year}
  {2016})}\BibitemShut {NoStop}%
\bibitem [{Note4()}]{Note4}%
  \BibitemOpen
  \bibinfo {note} {We note that the exact entropy function $w(\omega )$ cannot
  be expressed in terms of the microscopic data of the model by
  straightforwardly generalizing the standard arguments valid for unitary
  dynamics~\cite {AlbaPNAS}: the non-unitary generator $\protect \hat {\protect
  \mathcal {T}}$ is non-diagonalizable~\cite {lerose2020influence}, which makes
  the formulation of a generalized Gibbs ensemble for the dual dynamics
  {\protect \it a priori} unclear. It could presumably be obtained by
  generalizing the calculation of Ref.~\cite {CalabreseFagotti08} to
  non-unitary evolution.}\BibitemShut {Stop}%
\bibitem [{Note5()}]{Note5}%
  \BibitemOpen
  \bibinfo {note} {Here we are considering initial states with a pair structure
  of quasiparticles, as usually arising from global quenches~\cite
  {Calabrese_2005}. We note that fine-tuned examples are known for which this
  is not the case~\cite {bertini2018entanglement,bastianello2018spreading}. In
  this case, a different behavior could be envisaged.}\BibitemShut {Stop}%
\bibitem [{\citenamefont {Bertini}\ \emph
  {et~al.}(2018{\natexlab{a}})\citenamefont {Bertini}, \citenamefont {Kos},\
  and\ \citenamefont {Prosen}}]{BertiniSFF}%
  \BibitemOpen
  \bibfield  {author} {\bibinfo {author} {\bibfnamefont {B.}~\bibnamefont
  {Bertini}}, \bibinfo {author} {\bibfnamefont {P.}~\bibnamefont {Kos}}, \ and\
  \bibinfo {author} {\bibfnamefont {T.~c.~v.}\ \bibnamefont {Prosen}},\ }\href
  {\doibase 10.1103/PhysRevLett.121.264101} {\bibfield  {journal} {\bibinfo
  {journal} {Phys. Rev. Lett.}\ }\textbf {\bibinfo {volume} {121}},\ \bibinfo
  {pages} {264101} (\bibinfo {year} {2018}{\natexlab{a}})}\BibitemShut
  {NoStop}%
\bibitem [{\citenamefont {Ippoliti}\ and\ \citenamefont
  {Khemani}(2021)}]{ippoliti2021postselection}%
  \BibitemOpen
  \bibfield  {author} {\bibinfo {author} {\bibfnamefont {M.}~\bibnamefont
  {Ippoliti}}\ and\ \bibinfo {author} {\bibfnamefont {V.}~\bibnamefont
  {Khemani}},\ }\href {\doibase 10.1103/PhysRevLett.126.060501} {\bibfield
  {journal} {\bibinfo  {journal} {Phys. Rev. Lett.}\ }\textbf {\bibinfo
  {volume} {126}},\ \bibinfo {pages} {060501} (\bibinfo {year}
  {2021})}\BibitemShut {NoStop}%
\bibitem [{\citenamefont {Lu}\ and\ \citenamefont
  {Grover}(2021)}]{grover2021spacetime}%
  \BibitemOpen
  \bibfield  {author} {\bibinfo {author} {\bibfnamefont {T.-C.}\ \bibnamefont
  {Lu}}\ and\ \bibinfo {author} {\bibfnamefont {T.}~\bibnamefont {Grover}},\
  }\href {\doibase 10.1103/PRXQuantum.2.040319} {\bibfield  {journal} {\bibinfo
   {journal} {PRX Quantum}\ }\textbf {\bibinfo {volume} {2}},\ \bibinfo {pages}
  {040319} (\bibinfo {year} {2021})}\BibitemShut {NoStop}%
\bibitem [{\citenamefont {Ippoliti}\ \emph {et~al.}(2021)\citenamefont
  {Ippoliti}, \citenamefont {Rakovszky},\ and\ \citenamefont
  {Khemani}}]{ippoliti2021fractal}%
  \BibitemOpen
  \bibfield  {author} {\bibinfo {author} {\bibfnamefont {M.}~\bibnamefont
  {Ippoliti}}, \bibinfo {author} {\bibfnamefont {T.}~\bibnamefont {Rakovszky}},
  \ and\ \bibinfo {author} {\bibfnamefont {V.}~\bibnamefont {Khemani}},\ }\href
  {https://arxiv.org/abs/2103.06873} {\bibfield  {journal} {\bibinfo  {journal}
  {arXiv preprint arXiv:2103.06873}\ } (\bibinfo {year} {2021})}\BibitemShut
  {NoStop}%
\bibitem [{\citenamefont {Chan}\ \emph {et~al.}(2021)\citenamefont {Chan},
  \citenamefont {De~Luca},\ and\ \citenamefont {Chalker}}]{chan2021lyapunov}%
  \BibitemOpen
  \bibfield  {author} {\bibinfo {author} {\bibfnamefont {A.}~\bibnamefont
  {Chan}}, \bibinfo {author} {\bibfnamefont {A.}~\bibnamefont {De~Luca}}, \
  and\ \bibinfo {author} {\bibfnamefont {J.~T.}\ \bibnamefont {Chalker}},\
  }\href {\doibase 10.1103/PhysRevResearch.3.023118} {\bibfield  {journal}
  {\bibinfo  {journal} {Phys. Rev. Research}\ }\textbf {\bibinfo {volume}
  {3}},\ \bibinfo {pages} {023118} (\bibinfo {year} {2021})}\BibitemShut
  {NoStop}%
\bibitem [{\citenamefont {Garratt}\ and\ \citenamefont
  {Chalker}(2021{\natexlab{a}})}]{garratt2021pairing}%
  \BibitemOpen
  \bibfield  {author} {\bibinfo {author} {\bibfnamefont {S.~J.}\ \bibnamefont
  {Garratt}}\ and\ \bibinfo {author} {\bibfnamefont {J.~T.}\ \bibnamefont
  {Chalker}},\ }\href {\doibase 10.1103/PhysRevX.11.021051} {\bibfield
  {journal} {\bibinfo  {journal} {Phys. Rev. X}\ }\textbf {\bibinfo {volume}
  {11}},\ \bibinfo {pages} {021051} (\bibinfo {year}
  {2021}{\natexlab{a}})}\BibitemShut {NoStop}%
\bibitem [{\citenamefont {Garratt}\ and\ \citenamefont
  {Chalker}(2021{\natexlab{b}})}]{garratt2021mbdl}%
  \BibitemOpen
  \bibfield  {author} {\bibinfo {author} {\bibfnamefont {S.~J.}\ \bibnamefont
  {Garratt}}\ and\ \bibinfo {author} {\bibfnamefont {J.~T.}\ \bibnamefont
  {Chalker}},\ }\href {\doibase 10.1103/PhysRevLett.127.026802} {\bibfield
  {journal} {\bibinfo  {journal} {Phys. Rev. Lett.}\ }\textbf {\bibinfo
  {volume} {127}},\ \bibinfo {pages} {026802} (\bibinfo {year}
  {2021}{\natexlab{b}})}\BibitemShut {NoStop}%
\bibitem [{\citenamefont {Mi}\ \emph {et~al.}(2022)\citenamefont {Mi},
  \citenamefont {Sonner}, \citenamefont {Niu}, \citenamefont {Lee},
  \citenamefont {Foxen}, \citenamefont {Acharya}, \citenamefont {Aleiner},
  \citenamefont {Andersen}, \citenamefont {Arute}, \citenamefont {Arya} \emph
  {et~al.}}]{mi2022noise}%
  \BibitemOpen
  \bibfield  {author} {\bibinfo {author} {\bibfnamefont {X.}~\bibnamefont
  {Mi}}, \bibinfo {author} {\bibfnamefont {M.}~\bibnamefont {Sonner}}, \bibinfo
  {author} {\bibfnamefont {M.~Y.}\ \bibnamefont {Niu}}, \bibinfo {author}
  {\bibfnamefont {K.~W.}\ \bibnamefont {Lee}}, \bibinfo {author} {\bibfnamefont
  {B.}~\bibnamefont {Foxen}}, \bibinfo {author} {\bibfnamefont
  {R.}~\bibnamefont {Acharya}}, \bibinfo {author} {\bibfnamefont
  {I.}~\bibnamefont {Aleiner}}, \bibinfo {author} {\bibfnamefont {T.~I.}\
  \bibnamefont {Andersen}}, \bibinfo {author} {\bibfnamefont {F.}~\bibnamefont
  {Arute}}, \bibinfo {author} {\bibfnamefont {K.}~\bibnamefont {Arya}},  \emph
  {et~al.},\ }\href@noop {} {\bibfield  {journal} {\bibinfo  {journal} {arXiv
  preprint arXiv:2204.11372}\ } (\bibinfo {year} {2022})}\BibitemShut {NoStop}%
\bibitem [{\citenamefont {Dubail}(2017)}]{Dubail_2017}%
  \BibitemOpen
  \bibfield  {author} {\bibinfo {author} {\bibfnamefont {J.}~\bibnamefont
  {Dubail}},\ }\href {\doibase 10.1088/1751-8121/aa6f38} {\bibfield  {journal}
  {\bibinfo  {journal} {Journal of Physics A: Mathematical and Theoretical}\
  }\textbf {\bibinfo {volume} {50}},\ \bibinfo {pages} {234001} (\bibinfo
  {year} {2017})}\BibitemShut {NoStop}%
\bibitem [{\citenamefont {Wang}\ and\ \citenamefont
  {Zhou}(2019)}]{wang2019barrier}%
  \BibitemOpen
  \bibfield  {author} {\bibinfo {author} {\bibfnamefont {H.}~\bibnamefont
  {Wang}}\ and\ \bibinfo {author} {\bibfnamefont {T.}~\bibnamefont {Zhou}},\
  }\href {https://link.springer.com/article/10.1007/JHEP12(2019)020#citeas}
  {\bibfield  {journal} {\bibinfo  {journal} {Journal of High Energy Physics}\
  }\textbf {\bibinfo {volume} {2019}},\ \bibinfo {pages} {1} (\bibinfo {year}
  {2019})}\BibitemShut {NoStop}%
\bibitem [{\citenamefont {Reid}\ and\ \citenamefont
  {Bertini}(2021)}]{reid2021barrier}%
  \BibitemOpen
  \bibfield  {author} {\bibinfo {author} {\bibfnamefont {I.}~\bibnamefont
  {Reid}}\ and\ \bibinfo {author} {\bibfnamefont {B.}~\bibnamefont {Bertini}},\
  }\href {\doibase 10.1103/PhysRevB.104.014301} {\bibfield  {journal} {\bibinfo
   {journal} {Phys. Rev. B}\ }\textbf {\bibinfo {volume} {104}},\ \bibinfo
  {pages} {014301} (\bibinfo {year} {2021})}\BibitemShut {NoStop}%
\bibitem [{\citenamefont {Fr{\'\i}as-P{\'e}rez}\ and\ \citenamefont
  {Ba{\~n}uls}(2022)}]{frias2022light}%
  \BibitemOpen
  \bibfield  {author} {\bibinfo {author} {\bibfnamefont {M.}~\bibnamefont
  {Fr{\'\i}as-P{\'e}rez}}\ and\ \bibinfo {author} {\bibfnamefont {M.~C.}\
  \bibnamefont {Ba{\~n}uls}},\ }\href@noop {} {\bibfield  {journal} {\bibinfo
  {journal} {arXiv preprint arXiv:2201.08402}\ } (\bibinfo {year}
  {2022})}\BibitemShut {NoStop}%
\bibitem [{\citenamefont {Alba}\ and\ \citenamefont
  {Calabrese}(2017)}]{AlbaPNAS}%
  \BibitemOpen
  \bibfield  {author} {\bibinfo {author} {\bibfnamefont {V.}~\bibnamefont
  {Alba}}\ and\ \bibinfo {author} {\bibfnamefont {P.}~\bibnamefont
  {Calabrese}},\ }\href {\doibase 10.1073/pnas.1703516114} {\bibfield
  {journal} {\bibinfo  {journal} {Proceedings of the National Academy of
  Sciences}\ }\textbf {\bibinfo {volume} {114}},\ \bibinfo {pages} {7947}
  (\bibinfo {year} {2017})},\ \Eprint
  {http://arxiv.org/abs/https://www.pnas.org/content/114/30/7947.full.pdf}
  {https://www.pnas.org/content/114/30/7947.full.pdf} \BibitemShut {NoStop}%
\bibitem [{\citenamefont {Fagotti}\ and\ \citenamefont
  {Calabrese}(2008)}]{CalabreseFagotti08}%
  \BibitemOpen
  \bibfield  {author} {\bibinfo {author} {\bibfnamefont {M.}~\bibnamefont
  {Fagotti}}\ and\ \bibinfo {author} {\bibfnamefont {P.}~\bibnamefont
  {Calabrese}},\ }\href {\doibase 10.1103/PhysRevA.78.010306} {\bibfield
  {journal} {\bibinfo  {journal} {Phys. Rev. A}\ }\textbf {\bibinfo {volume}
  {78}},\ \bibinfo {pages} {010306} (\bibinfo {year} {2008})}\BibitemShut
  {NoStop}%
\bibitem [{\citenamefont {Bertini}\ \emph
  {et~al.}(2018{\natexlab{b}})\citenamefont {Bertini}, \citenamefont
  {Tartaglia},\ and\ \citenamefont {Calabrese}}]{bertini2018entanglement}%
  \BibitemOpen
  \bibfield  {author} {\bibinfo {author} {\bibfnamefont {B.}~\bibnamefont
  {Bertini}}, \bibinfo {author} {\bibfnamefont {E.}~\bibnamefont {Tartaglia}},
  \ and\ \bibinfo {author} {\bibfnamefont {P.}~\bibnamefont {Calabrese}},\
  }\href {\doibase 10.1088/1742-5468/aac73f} {\bibfield  {journal} {\bibinfo
  {journal} {J. Stat. Mech.}\ }\textbf {\bibinfo {volume} {2018}},\ \bibinfo
  {pages} {063104} (\bibinfo {year} {2018}{\natexlab{b}})}\BibitemShut
  {NoStop}%
\bibitem [{\citenamefont {Bastianello}\ and\ \citenamefont
  {Calabrese}(2018)}]{bastianello2018spreading}%
  \BibitemOpen
  \bibfield  {author} {\bibinfo {author} {\bibfnamefont {A.}~\bibnamefont
  {Bastianello}}\ and\ \bibinfo {author} {\bibfnamefont {P.}~\bibnamefont
  {Calabrese}},\ }\href {\doibase 10.21468/SciPostPhys.5.4.033} {\bibfield
  {journal} {\bibinfo  {journal} {SciPost Phys.}\ }\textbf {\bibinfo {volume}
  {5}},\ \bibinfo {pages} {33} (\bibinfo {year} {2018})}\BibitemShut {NoStop}%
\end{thebibliography}%

\end{document}